\def\nn#1 #2{#1, #2.}				% Name with 1 initial
\def\nnn#1 #2 #3{#1, #2. #3.}			% Name with 2 initials
\def\nnnn#1 #2 #3 #4{#1, #2. #3. #4.}		% Name with 3 initials
\def\rfprep#1;#2;#3 {{\par\rn#1 #2, preprint #3\par}}
\def\rn{}
\def\rf#1;#2;#3;#4;#5 {\par\rn#1 #2, {\it #3}, {\bf #4}, #5\par}
\def\rfbook#1;#2;#3;#4;#5 {{\frenchspacing\par\rn#1 #2, {\it #3} (#4: #5)\par}}
\def\rfproc#1;#2;#3;#4;#5;#6 {{\frenchspacing\par\rn#1 #2, in {\it #3}, ed. #4 (#5: #6)\par}}
\def\dualand{ \&\hbox{ }}				% Lower case "and" already in use.
\def\multiand{ \&\hbox{ }}				% Lower case "and" already in use.

\def\etal{{\frenchspacing\it et al.}}
\def\ie{{\frenchspacing\it i.e.}}
\def\eg{{\frenchspacing\it e.g.}}
\def\etc{{\frenchspacing\it etc.}}
\def\cf{{\frenchspacing cf.}}
\def\Cf{{\frenchspacing Cf.}}

\def\beq#1{\begin{equation}\label{#1}}
\def\eeq{\end{equation}}
\def\beqa#1{\begin{eqnarray}\label{#1}}
\def\eeqa{\end{eqnarray}}
\def\eq#1{equation~(\ref{#1})}

\def\fig#1{Figure~1}
\def\Fig#1{Figure~1}

\def\sec#1{Section~\ref{#1}}

\def\spose#1{\hbox to 0pt{#1\hss}}
\def\simlt{\mathrel{\spose{\lower 3pt\hbox{$\mathchar"218$}}
     \raise 2.0pt\hbox{$\mathchar"13C$}}}
\def\simgt{\mathrel{\spose{\lower 3pt\hbox{$\mathchar"218$}}
     \raise 2.0pt\hbox{$\mathchar"13E$}}}
\def\simpropto{\mathrel{\spose{\lower 3pt\hbox{$\mathchar"218$}}
     \raise 2.0pt\hbox{$\propto$}}}

\def\psiket{|\psi\rangle}
\def\up{|\!\!\uparrow\rangle}
\def\down{|\!\!\downarrow\rangle}
\def\smileface{\raisebox{-2pt}{$\ddot\smile$}}
\def\frownface{\raisebox{-2pt}{$\ddot\frown$}}
\def\deadface{\raisebox{0pt}{${\times\times\atop\frown}$}}
\def\noobs{|\,\ddot{\_}\,\rangle}
\def\noobs{|\,\ddot{\raisebox{-2pt}{-}}\,\rangle}
\def\upobs{|\smileface\rangle}
\def\deadobs{|\deadface\rangle}

\def\downobs{|\frownface\rangle}
\def\tensormult{\otimes} % For now

\def\r{{\bf r}}
\def\p{{\bf p}}
\def\ignore#{}

\def\ed{\end{document}}

\documentstyle[prd,twocolumn,aps]{revtex}
\begin{document}
\twocolumn[\hsize\textwidth\columnwidth\hsize\csname@twocolumnfalse\endcsname

%%%%%%%%%%%%%%%%%%%%%%%%%%%%%
\title{THE INTERPRETATION OF QUANTUM MECHANICS:\\
MANY WORLDS OR MANY WORDS?}

\author{Max Tegmark}

\address{Institute for Advanced Study, Princeton, NJ 08540; max@ias.edu}

\date{September 15, 1997}

\maketitle
%%%%%%%%%%%%%%%%%%%%%%%%%%%%%%%%%%%%%%%%%%%%%%%%%%%%%%%%%%%%%%%%%%%%%%%%%%%%%%%%%%%%%%%
%\noindent
%\begin{center}
%
%\vskip0.9truecm
%{\bf
%THE INTERPRETATION OF QUANTUM MECHANICS:\\
%MANY WORLDS OR MANY WORDS?
%}
%
%\vskip 0.5truecm
%
%Max Tegmark
%
%\smallskip
%
%{\footnotesize Institute for Advanced Study, Princeton, NJ 08540; max@ias.edu}
%
%\smallskip
%\vskip 0.2truecm
%
%\end{center}
%%%%%%%%%%%%%%%%%%%%%%%%%%%%%%%%%%%%%%%%%%%%%%%%%%%%%%%%%%%%%%%%%%%%%%%%%%%%%%%%%%%%%%%

\begin{abstract}
As cutting-edge experiments display ever more 
extreme forms of non-classical behavior, 
the prevailing view on the interpretation of 
quantum mechanics appears to be gradually 
changing. A (highly unscientific) poll taken 
at the 1997 UMBC quantum mechanics 
workshop gave the once all-dominant Copenhagen 
interpretation less than half of the votes. 
The Many Worlds interpretation (MWI) scored second, 
comfortably ahead of the Consistent 
Histories and Bohm interpretations. It is argued 
that since all the above-mentioned 
approaches to nonrelativistic quantum mechanics 
give identical cookbook prescriptions for 
how to calculate things in practice, 
practical-minded experimentalists, who have 
traditionally adopted the ``shut-up-and-calculate 
interpretation'', typically show little 
interest in whether cozy classical concepts are in fact real 
in some untestable metaphysical sense 
or merely the way we subjectively perceive a 
mathematically simpler world where the 
Schr\"odinger equation describes everything  --- and 
that they are therefore becoming less 
bothered by a profusion of worlds than by a 
profusion of words.

Common objections to the MWI are discussed.
It is argued that when environment-induced 
decoherence is taken into account, 
the experimental predictions of
the MWI are identical to those of the Copenhagen interpretation
except for an experiment involving a Byzantine form of 
``quantum suicide''. This makes the choice between them purely a matter
of taste, roughly equivalent to whether one believes 
mathematical language or human language to be more fundamental.

\end{abstract}
\bigskip
\bigskip

] % Must end \twocolumn command here,  or disaster occurs, for bizarre reasons.
%%%%%%%%%%%%%%%%%%%%%%%%%%%%%%%%%%%%%%%%%%%%%%

%%%%%%%%%%%%%%%%%%%%%%%%%%%%%%%%%%%%%%%%%%%%%%
\makeatletter
\global\@specialpagefalse
\def\@oddfoot{
\ifnum\c@page>1
  \reset@font\rm\hfill \thepage\hfill
\fi
\ifnum\c@page=1
  {\sl
%Available from
%h t t p://www.sns.ias.edu/$\tilde{~}$max/everett.html\hfill\\
To appear in ``Fundamental Problems in Quantum Theory'', 
{\frenchspacing eds. M. H. Rubin \& Y. H. Shih.}\hfill\\
}
\fi
} \let\@evenfoot\@oddfoot
\makeatother

\section{INTRODUCTION}

At the quantum mechanics workshop to which these 
proceedings are dedicated, held in August 1997 at UMBC, 
%the University of Maryland, 
the participants were 
polled as to their preferred interpretation of quantum 
mechanics. The results are shown in Table 1.
\begin{center}
\begin{tabular}{|l|r|} 
\hline
Interpretation&Votes\\
\hline
Copenhagen&13\\
Many Worlds&8\\
Bohm&4\\
Consistent Histories&4\\
Modified dynamics (GRW/DRM)&1\\
%MODIFIED HEISENBERG EQ (GRW/DRM)&0\\
%MODIFIED SCHROEDINGER EQ (GRW/DRM)&0\\
None of the above/undecided&18\\
\hline
\end{tabular}
\end{center}
Although the poll was highly informal and unscientific 
(several people voted more than once, many abstained, etc), 
it nonetheless indicated a rather striking shift in opinion 
compared to the old days when the Copenhagen interpretation 
reigned supreme. Perhaps most striking of all is that the 
Many Worlds interpretation (MWI), proposed by Everett in 
1957 \cite{Everett,EverettBook,Wheeler} but virtually 
unnoticed for about a decade \cite{Cooper,DeWitt},
has survived 25 years of 
fierce criticism and occasional ridicule to become 
the number one challenger to the leading orthodoxy,
ahead of the Bohm \cite{Bohm}, 
Consistent Histories \cite{ConsHist} and GRW \cite{GRW} 
interpretations.
Why has this happened? The purpose of the present 
paper is to briefly summarize the appeal of the MWI 
in the light of recent experimental and theoretical 
progress, and why much of the traditional criticism 
of it is being brushed aside.

\section{The MWI: what it is and what it isn't}

Much of the old criticism of the MWI was based on 
confusion as to what it meant. Here we grant Everett 
the final say in how the MWI is defined,  since he did 
after all invent it \cite{Everett}, and take it to consist of 
the following postulate alone:

\begin{itemize}
\item {\bf EVERETT POSTULATE:}\\
{\it All isolated systems evolve according to the Schr\"odinger equation
${d\over dt}\psiket=-{i\over\hbar}H\psiket.$}
%${d\over dt}\psiket={i\over\hbar}H\psiket.$}
\end{itemize}
Although this postulate sounds rather innocent, it has far-reaching 
implications:
\begin{enumerate} 
\item {\bf Corollary 1:} the entire Universe evolves 
according to the Schr\"odinger equation, since it is 
by definition an isolated system. 
\item {\bf Corollary 2:} there can be no definite outcome 
of quantum measurements (wavefunction collapse), since 
this would violate the Everett postulate.
\end{enumerate}
Because of corollary 1, ``universally valid quantum mechanics'' 
is often used as a synonym for the MWI. What is to be 
considered ``classical'' is therefore {\it not} specified axiomatically 
(put in by hand) in the MWI --- rather, it can be derived 
from the Hamiltonian dynamics as described in
\sec{DecoherenceSec}, by computing 
decoherence rates.

How does corollary 2 follow? 
Consider a measurement of a spin 1/2 system (a silver atom, say)
where the states 
``up'' and ``down''
along the $z$ axis are denoted $\up$ and $\down$. 
Assuming that the observer will get happy if she measures spin up, 
we let $\noobs$, $\upobs$ and $\downobs$ denote the states of 
the observer before the measurement, after perceiving 
spin up and after perceiving spin down, respectively.
If the measurement is to be described by a unitary 
Schr\"odinger time evolution operator $U=e^{-iH\tau/\hbar}$ applied to the
total system, then $U$ must clearly satisfy  
\beq{Ueq}
U\up\tensormult\noobs = \up\tensormult\upobs
\quad\hbox{and}\quad
U\down\tensormult\noobs = \down\tensormult\downobs.
\eeq
%Thus if the atom is originally in a superposition\\
Therefore if the atom is originally in a superposition
\hbox{$\alpha\up+\beta\down$}, then the Everett postulate implies that
the state resulting after the observer has interacted
with the atom is 
\beq{SplitEq}
U(\alpha\up+\beta\down)\tensormult\noobs = 
\alpha\up\tensormult\upobs + \beta\down\tensormult\downobs.
\eeq
In other words, the outcome is not $\up\tensormult\upobs$
or $\down\tensormult\downobs$ with some probabilities, merely 
these two states in superposition.
Very few physicists have actually read Everett's book 
(printed in \cite{EverettBook}), 
which has lead to a common misconception that it 
contains a second postulate along the following lines:
\begin{itemize}
\item What Everett does {\bf NOT} postulate:\\
{\it At certain magic instances, the the world undergoes
some sort of metaphysical ``split'' into 
two branches that subsequently never interact.}
\end{itemize}
This is not only a misrepresentation of the MWI, but also inconsistent 
with the Everett postulate, since the subsequent time evolution could in 
principle make the two terms in 
\eq{SplitEq} interfere. 
According to the MWI, there is, was and always will be
only one wavefunction, and 
only decoherence calculations, not postulates, 
can tell us when it is a good approximation to treat two terms 
as non-interacting.

\section{Common criticism of the MWI}

\subsection{``It doesn't explain why we perceive randomness"}
\label{ProbabilitySec}

Everett's brilliant insight was that the MWI 
{\it does} explain why we perceive
randomness even though the Schr\"odinger equation itself is competely
causal. To avoid linguistic confusion, it is crucial that we distinguish
between \cite{toe}
\begin{itemize}
\item the {\it outside view} of the world (the way a mathematical
thinks of it, \ie, as an  evolving wavefunction), and 
\item the {\it inside view}, the way it is perceived from the subjective
frog perspective of an observer in it. 
\end{itemize}
$\upobs$ and $\downobs$ have by definition perceived two opposite measurement outcomes
from their inside views, but share the same memory of being in the
state $\noobs$ moments earlier. Thus $\downobs$ describes an observer
who remembers performing a spin measurement and observing
the definite outcome $\down$. Suppose she measures the $z$-spin of
$n$ independent atoms that all have spin up in the $x$-direction initially,
\ie, $\alpha=\beta=1/\sqrt{2}$. The final state corresponding to \eq{SplitEq}
will then contain $2^n$ terms of equal weight, a typical term  
corresponding to a seemingly random sequence of ups and downs, 
of the form 
\beq{GenericTermEq}
2^{-n/2}
|\!\downarrow\downarrow\uparrow\downarrow\uparrow
\uparrow\uparrow\downarrow\downarrow\uparrow\cdots\rangle
\tensormult
|\frownface\frownface\smileface\frownface\smileface
\smileface\smileface\frownface\frownface\smileface\cdots\rangle.
\eeq
Thus the perceived inside view of what happened according to 
an observer described by a {\it typical} element of the final superposition
is a seemingly random sequence of ups and downs, behaving as if generated 
though a random process with probabilities $p=\alpha^2=\beta^2=0.5$ for each outcome.
This can be made more formal if we replace $``\frownface"$ by $``0"$, replace
$``\smileface"$ by ``1'', and place a decimal point in front of it all.
Then the above observer state $|\frownface\frownface\smileface\frownface\smileface
\smileface\smileface\frownface\frownface\smileface\cdots\rangle
= |.0010111001...\rangle$, and we see that in the limit $n\to\infty$, 
each observer state corresponds to a real number on the unit interval 
(written in binary). According to Borel's theorem on normal numbers
\cite{Borel,Chung},
almost all 
(all except for a set of Borel measure zero) real numbers
between zero and one have a fraction $0.5$ of their decimals
being ``1'', so in the same sense, 
almost all terms in our wavefunction describe observers that
have perceived the conventional quantum probability rules to hold.
It is in this sense that the MWI predicts apparent randomness from
the inside viewpoint while maintaining strict causality from the outside 
viewpoint.\footnote{It is interesting to note that Borel's 1909 theorem 
made a strong impression on many mathematicians of the time, some of whom
had viewed the entire probability concept with a certain suspicion,
since they were now confronted with a 
theorem in the heart of classical mathematics
which could be reinterpreted in terms of probabilities \cite{Chung}.
Borel would undoubtedly have been interested to know that his work showed 
the emergence of a probability-like concept ``out of the blue'' 
not only in in mathematics, but in physics as well.
}
For a clear and pedagogical generalization to the general case with unequal 
probabilities, see \cite{Everett,EverettBook}.

\subsection{``It doesn't explain why we don't perceive weird superpositions"}
\label{DecoherenceSec}

That's right! The Everett postulate doesn't! Since the state corresponding to 
a superposition of a pencil lying in two macroscopically 
different positions on a table-top is a perfectly permissible quantum
state in the MWI, why do we never perceive such states? Indeed, if we were
to balance a pencil exactly on its tip, it would by symmetry 
fall down in a superposition of all directions (a calculation shows that this takes
about 30 seconds), thereby creating such a macrosuperposition state.
The inability to answer this question was originally a serious weakness of the MWI, which can 
equivalently be phrased as follows: why is the position representation so special?
Why do we perceive macroscopic objects in approximate eigenstates of the position 
operator $\r$ and the momentum operator $\p$ but 
never in approximate eigenstates
of other Hermitian operators such as $\r+\p$?
The answer to this important question was provided by the realization 
that environment-induced decoherence
rapidly destroys macrosuperpositions as far as the inside view is 
concerned, but this was explicitly pointed out only
in the 70's \cite{Zeh70} and 80's \cite{Zurek81}, more than a decade after 
Everett's original work. 
This elegant mechanism is 
now well-understood and rather uncontroversial \cite{Zurek91}, and the
interested reader is referred to \cite{Omnes97} and a recent book on decoherence 
\cite{Giulini96} for details.
Essentially, the position basis gets singled out by the dynamics 
because the field equations of physics are local in this basis, 
not in any other basis. 

Historically, the collapse postulate was introduced to suppress 
the off-diagonal density matrix elements elements corresponding 
to strange macrosuperpositions ({\cf} \cite{Neumann}). 
However, many physicists have shared Gottfried's view that
``the reduction [collapse] postulate is an ugly scar on what would
be  beautiful theory if it could be removed'' \cite{Gottfried},
since it is not accompanied by any equation 
specifying {\it when} collapse occurs ({\it when} the Everett postulate is 
violated). 
% This motivated GRW  - BUT I'M SHORT ON SPACE!
The subsequent discovery of decoherence provided precisely
such an explicit mechanism for suppression of off-diagonal elements, which is 
essentially indistinguishable from the effect of a postulated
Copenhagen wavefunction collapse from an observational (inside)
point of view ({\eg} \cite{collapse}). Since 
this eliminates arguably the main motivation for the  
collapse postulate, it may be a principal reason for the 
increasing popularity of the MWI.

\subsection{``It's too weird for me"}

The reader must choose between two tenable but diametrically opposite
paradigms regarding physical reality and the status of mathematics:
\begin{itemize}
\item {\bf PARADIGM 1:} The outside view (the mathematical structure) is physically 
real, and the inside view and all the human language we use
to describe it is merely a useful approximation for describing
our subjective perceptions.
\item {\bf PARADIGM 2:} The subjectively perceived inside view is physically real, 
and the outside view and all its mathematical language is 
merely a useful approximation.
\end{itemize}
What is more basic --- the inside view or the outside view?
What is more basic --- human language or mathematical language?
Note that in case 1, which might be termed the {\it Platonic paradigm}, 
all of physics is ultimately a mathematics problem, since 
an infinitely intelligent mathematician given the equations of
the Universe could in principle
{\it compute} the inside view, {\ie}, 
compute what self-aware observers the Universe would contain, 
what they would perceive, and what language they would 
invent to describe their perceptions to one another. Thus in the
Platonic paradigm, 
the axioms of an ultimate ``Theory of Everything" would be purely
mathematical axioms, since axioms or postulates in English 
regarding interpretation would be derivable and thus redundant.
In paradigm 2, on the other hand, there can never be a 
``Theory of Everything'', since one is ultimately just explaining
certain verbal statements by other verbal statements ---
this is known as the infinite regress problem (\eg, \cite{Nozick}).

The reader who prefers the Platonic paradigm should find the MWI natural, whereas
the reader leaning towards paradigm 2 probably prefers 
the Copenhagen interpretation.
A person objecting that the MWI is ``too weird'' 
is essentially saying that the inside and 
outside views are extremely different, 
the latter being ``weird'', 
and therefore prefers paradigm 2.
In the Platonic paradigm, there is of course no reason whatsoever to expect the
inside view to resemble the outside view, so one expects the correct theory
to seem weird.
One reason why theorists are becoming increasingly positive to the MWI 
is probably that past theoretical breakthroughs have shown that the
outside view really {\it is} very different from the inside view.
For instance, a prevalent modern view of quantum field theory is 
(\eg, \cite{Nielsen,Weinberg})
that the standard model is merely an effective theory, a low-energy limit of 
a yet to be discovered theory that is 
even more removed from our cozy classical concepts (perhaps
involving superstrings in 26 dimensions, say). 
General Relativity has already
introduced quite a gap between the outside view (fields obeying covariant 
partial differential equations on a 4-dimensional manifold) 
and the inside view (where we always perceive
spacetime as locally Minowski, and our perceptions depend not only on 
where we are but also on how fast we are moving).

One reason why experimentalists are becoming increasingly positive to the MWI 
is probably that they have recently produced so many ``weird''
(but perfectly repeatable) experimental results (Bell inequality violations
with kilometer baselines \cite{Tapster}, 
molecule interferometry \cite{Pritchard}, vorticity 
quantization in a macroscopically large amount of liquid Helium \cite{Schwab}, 
\etc),
and therefore simply accept that the world is a weirder place than we thought
it was and get on with their calculations.

\subsection{``Many words" objections}

The questions addressed in Sections~\ref{ProbabilitySec}
and~\ref{DecoherenceSec} are in the author's opinion 
quite profound, and were answered thanks to the 
ingenuity of Everett and the discoverers
of decoherence, respectively. However, there are also a 
number of questions/objections
that in the author's opinion belong in the category 
``many words'', being issues of semantics rather than physics.
When discussing the MWI, it is of course within the context of the
Platonic paradigm described above, paradigm 1, 
where equations are ultimately more fundamental than
words. Since human language is merely something that certain
observers have invented to describe their subjective 
perceptions, many words describe concepts
that by necessity are just useful approximations 
({\cf} \cite{Page95}).
We know that the classical concept of gas pressure is merely 
an approximation that breaks down if we consider atomic 
scales, and in the Platonic paradigm, 
we should not be surprised if we find
that other traditional concepts 
(\eg, that of physical probability, and indeed the entire
notion of a classical world) also turn out to be merely convenient 
approximations.

As an example of a ``many words'' objection, let us consider the rather 
subtle claim that the MWI does not justify
the use of the word ``probability'' \cite{Albert}.
When our observer is described by the state $\noobs$
before measuring her atom, there is no aspect of the measurement
outcome of which she has epistemological uncertainty 
(lack of knowledge): she simply knows that with 100\% certainty,
she will end up in a superposition of $\upobs$ and $\downobs$.
After the measurement, there is still no epistemological 
uncertainty, since both $\upobs$ and $\downobs$ know what they have
measured. For those who feel that the word probability 
should only be used when there is true lack of knowledge, 
probabilities can readily be introduced by 
performing the experiment while the observer is sleeping,
and placing her bed in one of two identical rooms depending on 
the outcome \cite{Vaidman}. On awakening, the observer described
by either of the two states in the superposition
can thus say
that she is in the first room with 50\% probability in the sense that
she has lack of knowledge as to where she is.
If there were $2^n$ identical rooms and $n$ measurements dictated
the room number in binary, then the observers in the final superposition
could compute probabilities for observing specific numbers of zeroes
and ones in their room number. Moreover, these 
could have been computed in advance of the experiment,
used as gambling odds, \etc, before the orthodox linguist would 
allow us to call them probabilities, which is why they are a useful concept
regardless of what we call them. 

Let us also consider a paper entitled 
``Against Many-Worlds Interpretations'' by 
Kent \cite{Kent90,Kent97}. Although most of its claims were 
subsequently shown to result from misconceptions \cite{Sakaguchi} 
(as to the definition of the MWI, as to the
mathematical distinction between ``measure'' [of a subset]
and ``norm'' [of a vector], \etc), it also contained 
a number of objections in the ``many words'' category.
In Section II.A, the author states that 
{\it ``one needs to define [...] the preferred basis [...] by an axiom.''}
According to what preconceived notion is this necessary, since 
decoherence can determine the preferred basis dynamically?
In the foreword to a 1997 version of this paper \cite{Kent97},
it is not only suggested that MWI adherents
{\it ``represent a relatively small minority''} and 
{\it ``tend to be working in other areas of physics''}
(both in apparent contradiction to the above-mentioned poll),
but also that they {\it ``tend to 
have non-standard views on the nature of scientific theories''}.
In our terminology, this ``objection'' presumable reflects the obvious
fact that MWI adherents subscribe to paradigm 1 rather than 2.
Moreover, Galileo once held ``non-standard'' views on 
the epicycle theory of planetary motion.

A large number of other objections have  been raised against the MWI,
tacitly based on some variant of paradigm 2. 
The opinion of this author is that 
{\it if paradigm 1 is adopted},  then there are 
{\it no} outstanding problems with the MWI when  decoherence is taken into
account (as discussed in \eg \cite{Giulini96,collapse,Zeh93}).

%\section{DOES THE MWI MAKE ANY DIFFERENCES IN PRACTICE?}
\section{IS THE MWI TESTABLE?}

\subsection{The ``shut-up-and-calculate'' recipe}

When comparing the contenders in Table 1, 
it is important to distinguish between their experimental
predictions and their philosophical interpretation.
When confronted with experimental questions, adherents of 
the first four will all agree on the following cookbook 
prescription for how to compute the right answer, which we
will term the 
``shut-up-and-calculate''\footnote{The author is indebted to Anupam Garg
for this phrase.} 
recipe:
\begin{quote}
{\it
Use the Schr\"odinger equation in all your calculations.
To compute the probability for what you personally 
will perceive in the end, simply 
convert to probabilities in the traditional way 
at the instant when you become mentally
aware of the outcome. In practice, you can 
convert to probabilities much earlier, as soon
as the superposition becomes ``macroscopic'',
and you can determine when this occurs by a standard
decoherence calculation.
}
\end{quote}
The fifth contender (a dynamical reduction mechanism such as
that proposed by Ghirardi, Rimini \& Weber) 
is the only one in the table to prescribe 
a different calculational recipe, since it modifies the 
Heisenberg equation of motion $\dot\rho=-i[H,\rho]/\hbar$
by adding an extra term \cite{GRW}.

\subsection{Quantum suicide}

The fact that the four most popular contenters in Table 1
have given identical
predictions for all experiments performed so far probably explains
why practical-minded physicists show so 
little interest in interpretational questions. Is there then
{\it any} experiment that could distinguish between say the 
MWI and the Copenhagen interpretation using currently 
available technology?
({\Cf} \cite{Deutsch,Lockwood}.) The author can only think of 
one: a form of quantum suicide in a spirit similar to 
so-called quantum roulette. %  (REF quantum roulette).
It requires quite a dedicated experimentalist, since it 
is amounts to an iterated and faster version of 
Schr\"odinger's cat experiment \cite{Erwin} --- with you as the cat. 

The apparatus is a ``quantum gun'' which each time its trigger
is pulled measures the $z$-spin of a 
particle in the state $(\up+\down)/\sqrt{2}$.
It is connected to a machine gun that fires a single bullet
if the result is ``down" and merely makes an audible click if the
result is ``up''.
The details of the trigger mechanism are irrelevant 
(an experiment with photons and a half-silvered mirror
would probably be cheaper to implement) as long as the timescale between the 
quantum bit generation and the actual firing is much shorter than 
that characteristic of human perception, say $10^{-2}$ seconds. 
The experimenter first places a sand bag in front of the 
gun and tells her assistant to pull the trigger ten times.
All contenders in Table 1 agree that the ``shut-up-and-calculate''
prescription applies here, and predict that she will hear a 
seemingly random sequence of shots and 
duds such as 
``bang-click-bang-bang-bang-click-click-bang-click-click.'' 
She now instructs her assistant to pull the trigger ten 
more times and places her head in front of the gun barrel.
This time the shut-up-and-calculate recipe is inapplicable, since 
probabilities have no meaning for an observer in the dead state
$\deadobs$, and the contenders will differ in their predictions.
In interpretations where there is an explicit non-unitary collapse, 
she will be either dead or alive after the first trigger event,
so she should expect to perceive perhaps a click or
two (if she is moderately lucky), then ``game over'', nothing at all.
In the MWI, on the other hand, the state after 
the first trigger event is 
\beq{SuicideEq}
U\>{1\over\sqrt{2}}\biggl(\up+\down\biggr)\tensormult\noobs 
= {1\over\sqrt{2}}\biggl(\up\tensormult\upobs+\down\tensormult\deadobs\biggr).
\eeq
Since there is exactly one observer having perceptions both
before and after the trigger event, and since it occurred too fast to 
notice, the MWI prediction is that $\noobs$ will
hear ``click'' with $100\%$ certainty. When her assistant has 
completed his unenviable assignment, she will have heard 
ten clicks, and concluded that collapse interpretations of 
quantum mechanics are ruled out at a confidence level 
of $1-0.5^n\approx 99.9\%$. If she wants to rule them out at 
``ten sigma'', she need merely increase $n$ by continuing 
the experiment a while longer.
Occasionally, to verify that the apparatus is working, she can move her head
away from the gun and suddenly hear it going off intermittently.
Note, however, that almost all terms in the final superposition
will have her assistant perceiving that he has killed his boss.

Many physicists would undoubtedly rejoice if an omniscient genie appeared at
their death bed, and as a reward for life-long curiosity granted them
the answer to a physics question of their choice. But would they be as 
happy if the genie forbade them from telling anybody else?
Perhaps the greatest irony of quantum mechanics is that if the MWI 
is correct, then the situation is quite analogous if 
once you feel ready to die, you repeatedly attempt quantum suicide:
you {\it will} experimentally convince 
yourself that the MWI is correct, but you can never convince anyone else!

\bigskip

The author wishes to thank David Albert, Orly Alter, Geoffrey Chew,
Ang\'elica de Oliveira-Costa,
Michael Gallis, Bill Poirier,
Svend Erik Rugh, Marlan Scully, Robert Spekkens, Lev Vaidman, 
John Wheeler and Wojciech Zurek (some of whom disagree passionately with the opinions 
expressed in the present paper!) for thought-provoking and 
entertaining discussions. This work was supported by Hubble Fellowship
{\#}HF-01084.01-96A, awarded by the 
Space Telescope Science Institute, which is operated by AURA, Inc. 
under NASA contract NAS5-26555.

%\end{document}

%\end{thebibliography}

\bigskip
\bigskip
\bigskip
\noindent
{\sl 
{\bf NOTE:}\\
This paper and a number of related ones are available online at
h t t p://www.sns.ias.edu/$\tilde{~}$max/everett.html\hfill\\
}

\end{document}